\documentclass[twocolumn, 11pt]{aastex61}

\usepackage{natbib}
\usepackage{graphicx}
\usepackage{bm}
\usepackage{amsmath}
\usepackage{amsfonts}
\usepackage{amssymb}

\newcommand{\unit}[1]{\ifmmode \:\mbox{\rm #1}\else \mbox{#1}\fi} 
\newcommand{\mamo}[1]{\mbox{$#1$}}

\def\sun{\hbox{$\odot$}}

\newcommand{\kpc}{\unit{kpc}}

\newcommand{\R}{\textbf{R}\ }

\newcommand{\msun}{\mamo{M_{\sun}}}

\received{14 Oct 2018}
\accepted{12 March 2019}
\revised{11 March 2019}

\shortauthors{Eadie \& Juri\'c}

\begin{document}

	\title{The cumulative mass profile of the Milky Way \\ as determined by globular cluster kinematics from Gaia DR2}

	\correspondingauthor{Gwendolyn Eadie}
	\email{eadieg@uw.edu}
		\author[0000-0003-3734-8177]{Gwendolyn Eadie}
	\affil{Department of Astronomy,	University of Washington, Seattle, WA, USA}
	\affil{eScience Institute, University of Washington, Seattle, WA, USA}
	\affil{DIRAC Institute, University of Washington, Seattle, WA, USA}
	
	\author{Mario Juri\'c}
	\email{mjuric@astro.washington.edu}
	\affil{Department of Astronomy,	University of Washington, Seattle, WA, USA}
	\affil{eScience Institute, University of Washington, Seattle, WA, USA}
	\affil{DIRAC Institute, University of Washington, Seattle, WA, USA}

\begin{abstract}
		
We present new mass estimates and cumulative mass profiles (CMPs) with Bayesian credible regions for the Milky Way (MW) Galaxy, given the kinematic data of globular clusters as provided by (1) the \emph{Gaia} DR2 collaboration and the HSTPROMO team, and (2) the new catalog in Vasiliev (2019). We use globular clusters beyond 15kpc to estimate the CMP of the MW, assuming a total gravitational potential model $\Phi(r) = \Phi_{\circ}r^{-\gamma}$, which approximates an NFW-type potential at large distances when $\gamma=0.5$. We compare the resulting CMPs given data sets (1) and (2), and find the results to be nearly identical. The median estimate for the total mass is $M_{200}=  0.70 \times 10^{12} \msun$ and the 50\% Bayesian credible interval is $(0.62, 0.81)\times10^{12}\msun$. However, because the Vasiliev catalog contains more complete data at large $r$, the MW total mass is  slightly more constrained by these data. In this work, we also supply instructions for how to create a CMP for the MW with Bayesian credible regions, given a model for $M(<r)$ and samples drawn from a posterior distribution. With the CMP, we can report median estimates and 50\% Bayesian credible regions for the MW mass within any distance (e.g., $M(r=25\kpc)= 0.26~(0.20, 0.36)\times10^{12}\msun$, $M(r=50\kpc)= 0.37~(0.29, 0.51) \times10^{12}\msun$, $M(r=100\kpc) = 0.53~(0.41, 0.74) \times10^{12}\msun$, etc.), making it easy to compare our results directly to other studies.

\end{abstract}


	\section{Introduction}\label{sec:intro}

	Since the \emph{Gaia} Data Release 2 (DR2) in April 2018, a number of studies have presented new estimates or lower bounds for the mass of the Milky Way (MW) Galaxy \citep[e.g.][]{2018GaiaDR2HelmiGCS, malhanibata2018arXiv, 2019PostiHelmiAA, 2018SohnApJ, Watkins2019ApJ, Vasiliev2019MNRAS}. Approaches have included maximum likelihood and Bayesian analyses, and have used data for kinematic tracers such as globular clusters (GCs) and stellar streams provided by DR2.
	
	These studies usually interpret results from the inferred gravitational potential profile or the circular velocity profile given an assumed model and the data. Results are often reported as a mass within a specific distance from the Galactic center, or as $M_{200}$ or the virial mass. While it is useful to have visualizations of the potential and circular velocity profiles in addition to individual estimates of the MW mass within specific distances, it would also be beneficial for studies to present a cumulative mass profile (CMP) of the MW with credible regions. A CMP  makes it straightforward to compare mass results reported within different Galactocentric distances, and it can also be compared to the CMP of MW-type galaxies from cosmological, hydrodynamical simulations. Moreover, CMPs resulting from different model assumptions can be compared and used to assess differences in model fits at different distances.
	
	In \cite{2015EHW} (Paper 1), \cite{eadie2016} (Paper 2), and \cite{ESH2017} (Paper 3), a hierarchical Bayesian method was developed for estimating the total mass and CMP of the MW. This Bayesian model was applied to the MW's GC population data, using the \citealt{1996harrisPaper} catalog (\citeyear{2010Harris} edition) and supplemented with proper motion measurements made by many other studies \citep[e.g.][see Paper 2 for a complete table]{majewski1993, zoccali2001, feltzing2002, dinescu2010, dinescu2013, fritz2015, rossi2015}.  
		
	In this paper, we provide the method to calculate a CMP given any model for the total gravitational potential and the posterior distribution of the model parameters acquired from a Bayesian analysis. As a motivating example, we employ a simple model for the total gravitational potential and CMP, and confront the model with \textit{Gaia} DR2 data of GC kinematics to arrive at a new mass estimate and CMP for the MW. For reasons outlined in Section~\ref{sec:methods}, we only use GCs that reside beyond 15kpc. This limits the sample considerably, and also increases the percentage of GCs without proper motion measurements.

	 The kinematic measurements of GCs released by the Gaia DR2 collaboration do not vastly improve the completeness of tracer data beyond 15kpc, but the HSTPROMO project has helped contribute valuable measurements for GCs at large distances \citep{2018SohnApJ}. Additionally, \cite{Vasiliev2019MNRAS} estimated the mean proper motion of 150 GCs using the Gaia DR2 dataset, thereby increasing the number of proper motion measurements at larger distances considerably\footnote{After the original submission of our manuscript, a similar catalog was released by \cite{Baumgardt2019MNRAS}.}. The catalog provided by Vasiliev is appealing not only because the data are more complete, but also because (1) all estimates of GC proper motions are measured in a consistent manner, and (2) the measurements agree very well with the available \emph{Gaia} and HSTPROMO observations of 20 distant GCs by \cite{2018SohnApJ}.
	
	Therefore, we estimate the mass and CMP of the MW using first the Gaia collaboration and the HSTPROMO project data, and then again using the catalog presented in \cite{Vasiliev2019MNRAS}. To obtain the posterior distribution of model parameters that will be used to determine the CMP, we use the hierarchical Bayesian method presented in Papers 1--3, with small adjustments in light of the results in \cite{EKH2018} (hereafter EKH). 
	
	The paper is organized as follows:
	\begin{itemize}
		\item Section~\ref{sec:methods} outlines the hierarchical Bayesian model from Papers 1-3 and the methods used in this study. In Section~\ref{sec:CMP}, we describe how to calculate the CMP for the Milky Way given a function $M(R<r)$, and given samples from the posterior distribution of model parameters. 
		\item Section~\ref{sec:data} discusses the details of and differences between the two data sets (Gaia DR2 + HSTPROMO, and Vasiliev).
		\item Section~\ref{sec:results} presents the Bayesian estimates of the MW's CMP, given each data set, and some discussion and interpretation of results.
		\item Section~\ref{sec:conclude} summarizes our findings and the impact on future studies.
	\end{itemize}

	\section{Method}\label{sec:methods}
	
	Specific details about our hierarchical Bayesian model and sampling methods for the posterior distribution can be found in Papers 1-3. Here, we provide a brief review for completeness.
	
	The method allows the user to supply (1) kinematic and position data of tracers (such as GCs or halo stars), (2) an analytic distribution function (DF) for the specific energies $\mathcal{E}$ and angular momenta $L$ of these tracers (based on a gravitational potential and tracer density profile), and (3) hyperprior distributions for the model parameters. 	The hierarchical Bayesian model includes incomplete and complete velocity data simulataneously, and also accounts for observational error.

	The observations of the kinematic tracers are the Galactocentric distance $r$, line-of-sight velocity $v_{\text{los}}$, and proper motions ($\mu_{\alpha}\cos{\delta}$, $\mu_\delta$), and their uncertainties. These measurements are assumed to be samples drawn from normal distributions centered on the tracers' true but unknown values with standard deviations equal to the measurement uncertainties. In other words, the true tracer velocity components and distances are treated as parameters. This measurement model defines the likelihood in the hierarchical Bayesian model. 
	
	The prior distribution on the parameters for the true values of $r$, $v_{\text{los}}, \mu_{\alpha}\cos{\delta}$, \text{and} $\mu_\delta$ is a distribution function (DF) of the specific energies $\mathcal{E}$ and angular momenta $L$, given a model for the total gravitational potential $\Phi(r)$ and density profile of the tracer population $\rho(r)$. In Papers 2 and 3, we used a simple power law profile for both the gravitational potential,
	\begin{equation}
	\Phi(r) = \Phi_{\circ}r^{-\gamma}, 
	\label{eq:potential}
	\end{equation}		
	and the density profile of the tracer population,
	\begin{equation}
	\rho \propto r^{-\alpha},
	\label{eq:density}
	\end{equation}
	where $\Phi_{\circ}, \gamma, \text{and~} \alpha$ are parameters. 
	At large distance, a spherically symmetric gravitational potential (equation~\ref{eq:potential}) is a reasonable approximation; recent studies regarding the velocity ellipsoid of the stellar halo have suggested the potential is more spherical than flattened \citep{2016WilliamsetalMNRAS, Wegg2019MNRAS}.
	
	Together, equations~\ref{eq:potential} and \ref{eq:density}, determine the analytic DF $f(\mathcal{E}, L)$ first presented in \cite{evans1997}. The DF also includes the constant velocity anisotropy parameter $\beta$ for the tracer population as a free parameter. The derivation of this DF is shown in \cite{evans1997}, with a condensed version shown in Paper 2 (although notations differ).

	Hyperprior distributions are also defined for the four model parameters: $\Phi_{\circ}, \gamma, \alpha, \text{and } \beta$. In Papers 2--3, the parameters $\Phi_{\circ}, \gamma, \text{and~} \beta$ were given uniform prior distributions with upper and lower bounds of $(1,200)$, $(0.3, 0.7)$, and $(-0.5, 1)$ respectively, while the prior distribution on $\alpha$ was a Gamma distribution determined by data not used in the analysis. 

	The posterior distribution is the probability distribution of model parameters given the likelihood, prior, hyperpriors, and the data. Instead of calculating the normalization constant in Bayes' theorem, samples from a distribution proportional to the posterior distribution are acquired. This is done by running multiple Markov chains in parallel, until they have reached a common, stationary distribution assumed to be proportional to the posterior distribution. Both visual inspection and statistical diagnostics are used to assess the mutual convergence of the chains.  The code, called \emph{GME}, was written in the \R Statistical Software environment (see Papers 1--3). 

	The hierarchical Bayesian model outlined above was used to estimate the total mass and CMP of the MW, given kinematic data for the MW GC population before \emph{Gaia} DR2 became available. More recently, EKH tested this method on tracer data from simulated galaxies made in the McMaster Unbiased Galaxy Simulations 2 (MUGS2).
	
	The study of MUGS2 simulated data was done as a strictly blind test. There were two main caveats to the study: (1) the tracer data from MUGS2 were globular cluster \emph{analogs} (i.e. star tracer particles instead of actual GCs), and (2) half of the galaxies created in the MUGS2 simulations did not follow the standard stellar-mass-to-halo-mass relation. Despite these caveats, the galaxies had the basic components of a bulge, disk, and dark matter halo, and it was therefore instructive to observe how well the single power law gravitational potential model could describe the overall system. 
	
	The results of the blind tests were mixed; the total mass ($M_{200}$) of some galaxies were estimated well within the 95\% Bayesian credible regions, while others were not. In particular, there was a tendency for the mass to be underestimated. 	EKH suggested a number of compounding reasons for the bias (e.g. the physical location of incomplete data and the spatial distribution of tracers in the simulation), but a main factor seemed to be the model for the total gravitational potential. The model had difficulty predicting both the inner and outer regions of the true CMPs for the simulated galaxies when tracers at all Galactocentric radii were included.

	Of the galaxies whose masses were poorly estimated, the posterior distribution was pushed towards and reached the edge of allowable parameter space. Posterior distributions that are at an extreme end of parameter space allowed by the prior distribution usually indicate a poor fit to the data, and can be a red flag that the results should be interpreted with caution. Moving forward, we keep this in mind as we apply the method of Paper 3 to the new Gaia data.
	
	EKH also showed that using only the outermost tracers from the simulated galaxies improved the mass estimates and increased the chance of containing the true mass profile within the Bayesian credible regions. These results echoed the sensitivity test in Paper 3 which also used a power law potential. In the latter case, the removal of inner GC data from the analysis caused a slight increase in the mass estimate for the MW. Thus, EKH recommended that in future studies, only outer kinematic tracers be used in the analysis when a single power law model for the gravitational potential is employed. For this study, we use GCs at $r>15\kpc$.

\begin{figure}
		\centering
		\includegraphics[scale=0.4]{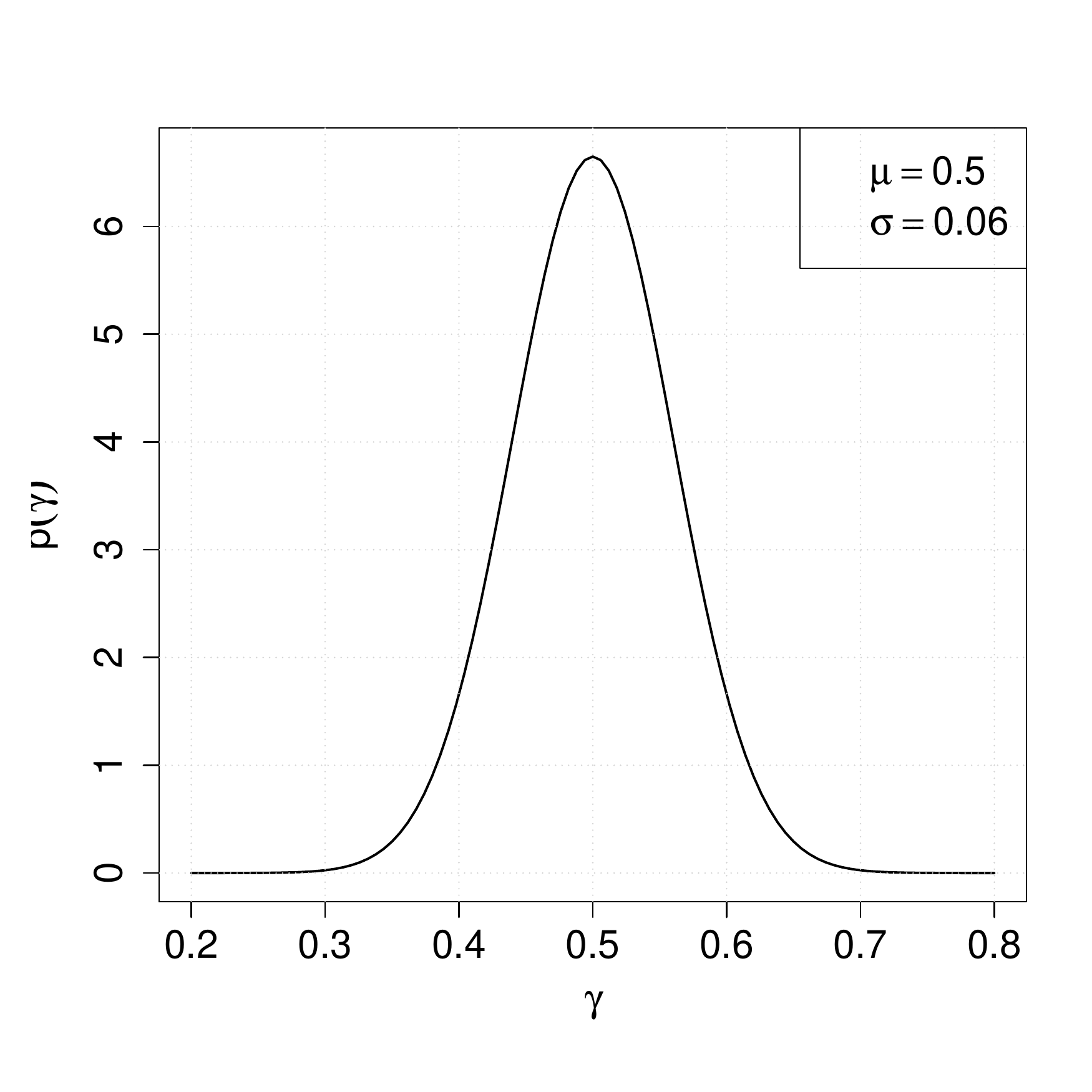}
		\caption{Prior distribution for $\gamma$, with mean 0.5 and standard 0.06.}\label{fig:priorgamma}
	\end{figure}

	EKH also found that using a uniform distribution as a prior on $\gamma$ may have been too broad. The uniform distibution was centered on 0.5, but also allowed values as low and as high as 0.3 and 0.7. When $\gamma=0.5$, Equation~\ref{eq:potential} approaches a Navarro-Frenk-White-type halo at large distance, and this is the prior information we intended to include. Thus, for this study, we replace the uniform distribution for $p(\gamma)$ with a normal distribution with mean $0.5$ and standard deviation $0.06$ (Figure~\ref{fig:priorgamma}).
	
\begin{figure}[h]
		\centering			
		\includegraphics[scale=0.4]{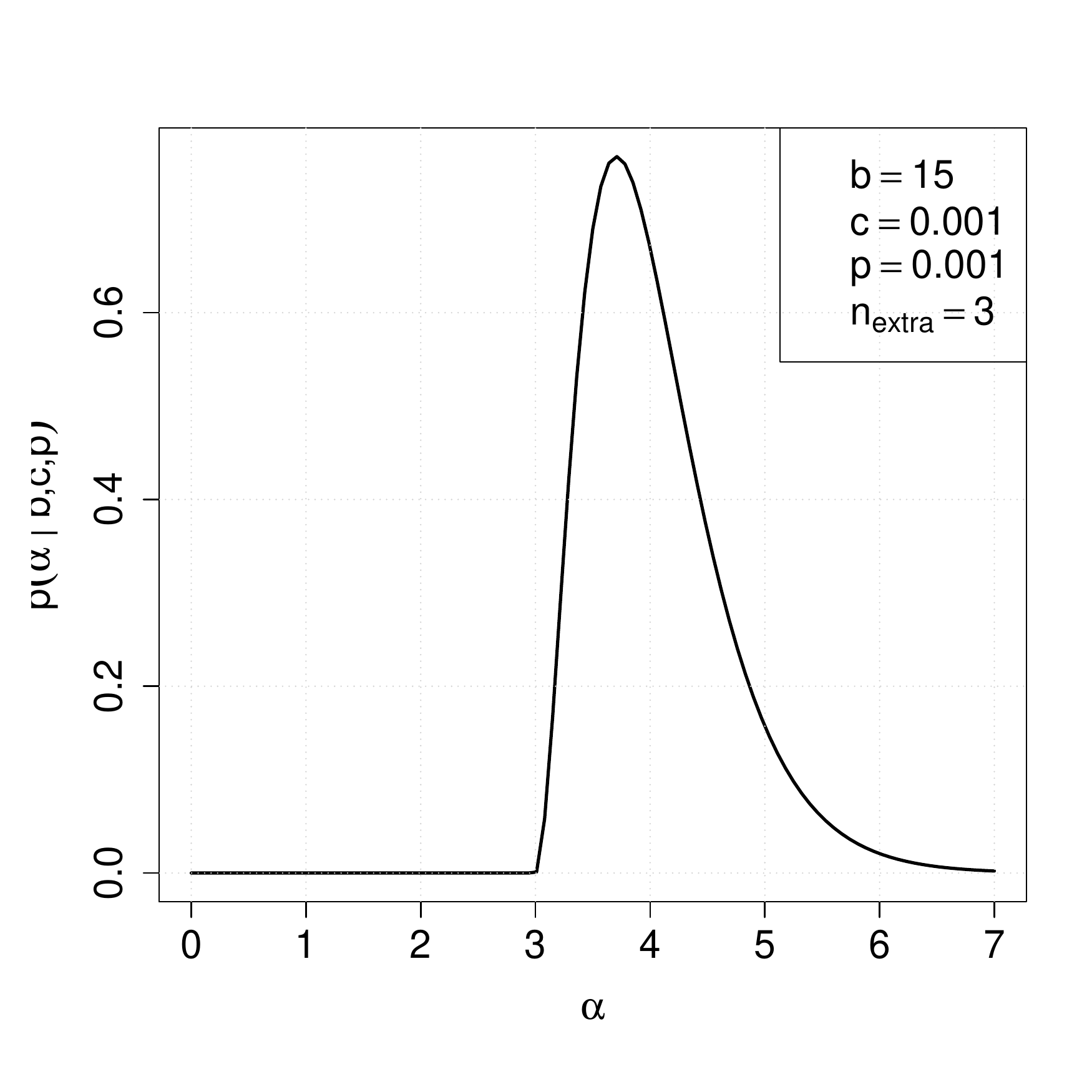}
		\caption{Prior distribution for $\alpha$, given the positions of AM 4, Ko 1, and Ko 2.}\label{fig:prioralpha}
		\end{figure}

	In this study, we continue to use a Gamma distribution for the $p(\alpha)$ with hyperparameters defined by GC data not used in the analysis because of their lack of $v_{\text{los}}$ and $\bar{\mu}$ measurements (Figure~\ref{fig:prioralpha}). Because we are excluding GCs within 15kpc, there are only three GCs rith $r>15\kpc$ that are lacking both line-of-sight and proper motion measurements: AM 4, Ko 1, and Ko 2.

		\subsection{Calculating the CMP with Bayesian credible regions}\label{sec:CMP}
	
	To calculate Bayesian credible regions for a CMP, given samples from the posterior distribution, perform the following steps:
	
	\begin{enumerate}
		\item Create a sequence of $n$ closely spaced $r$ values for the desired range from the Galactic center.
		
		\item At each $r_i$ for $i$ in $(1, ... , n)$, calculate $M(<r_i)$ for every sample from the posterior distribution. For each $r_i$ value, you should have $m$ values of $M(<r_i)$, where $m$ is the number of samples from the posterior distribution (e.g. the length of the Markov chain).
		
		\item At each $r_i$, sort the $m$ values of $M(<r_i)$, and calculate the credible intervals of interest (e.g. inner 50\%, 75\%, and 95\%). 
		
	\end{enumerate}
	
	 In our study, the total gravitational potential assumed in Equation~\ref{eq:potential} leads to a CMP of the form
	\begin{equation}
	M(<r) = \frac{\gamma\Phi_o}{G}\left(\frac{r}{\text{kpc}}\right)^{1-\gamma}
	\label{eq:Mr}
	\end{equation}
	\citep{Deason2012ApJ}. Thus, for a Markov chain of length $m$, each  $(\Phi_{\circ, j}, \gamma_j)$ pair in the chain is used to calculate the mass within a given radius $r_i$. This results in a vector of masses within $r_i$, which are sorted and used to calculate the marginal distribution of the mass within $r_i$. We use $n=100$ values logarithmically spaced from $r_1=0.1\kpc$ to $r_{100}=200\kpc$. Our chains are of length 90000, with effective sample sizes of $>2000$ for all parameters.

	\section{Data}\label{sec:data}
		
	As part of Gaia DR2, \cite{2018GaiaDR2HelmiGCS} published kinematic data for 75 GCs in the Milky Way, including proper motions, positions, distances, and line-of-sight velocities\footnote{\url{https://www.astro.rug.nl/~ahelmi/research/dr2-dggc/}}. We use these data to replace the relevant GC measurements in the \citealt{1996harrisPaper} catalog (\citeyear{2010Harris} edition), with some exceptions.
	
	The distance measurements to GCs as measured by Gaia show a systematic difference that should improve with the next data release \citep{2018GaiaDR2HelmiGCS}. In the meantime, we follow the \cite{2018GaiaDR2HelmiGCS} guidelines and continue to use the distances provided in \cite{2010Harris}. The Gaia data are also not entirely complete--- only 57 of the 75 GCs have line-of-sight velocity measurements. For those that are missing $v_{los}$ measurements, we use the ones listed in \cite{2010Harris} when available.
	
	The HSTPROMO team recently provided proper motions for 20 GCs at larger distances \citep{2018SohnApJ}. Four GCs measured by HSTPROMO  are also measured by Gaia (NGC 362, NGC 2298, NGC 2808, NGC 3201), and in these cases we use the DR2 data. In all other cases, we let the HSTPROMO measurements of GCs supersede any previous measurements in the literature. We then supplement other missing proper motions measurements with those reported in other studies \citep{majewski1993, zoccali2001, dinescu2010, dinescu2013, delaFuenteMarcos2015AA, rossi2015}.
		\begin{figure}
		\centering
		\includegraphics[scale=0.4, trim=3cm 3cm 3cm 4cm]{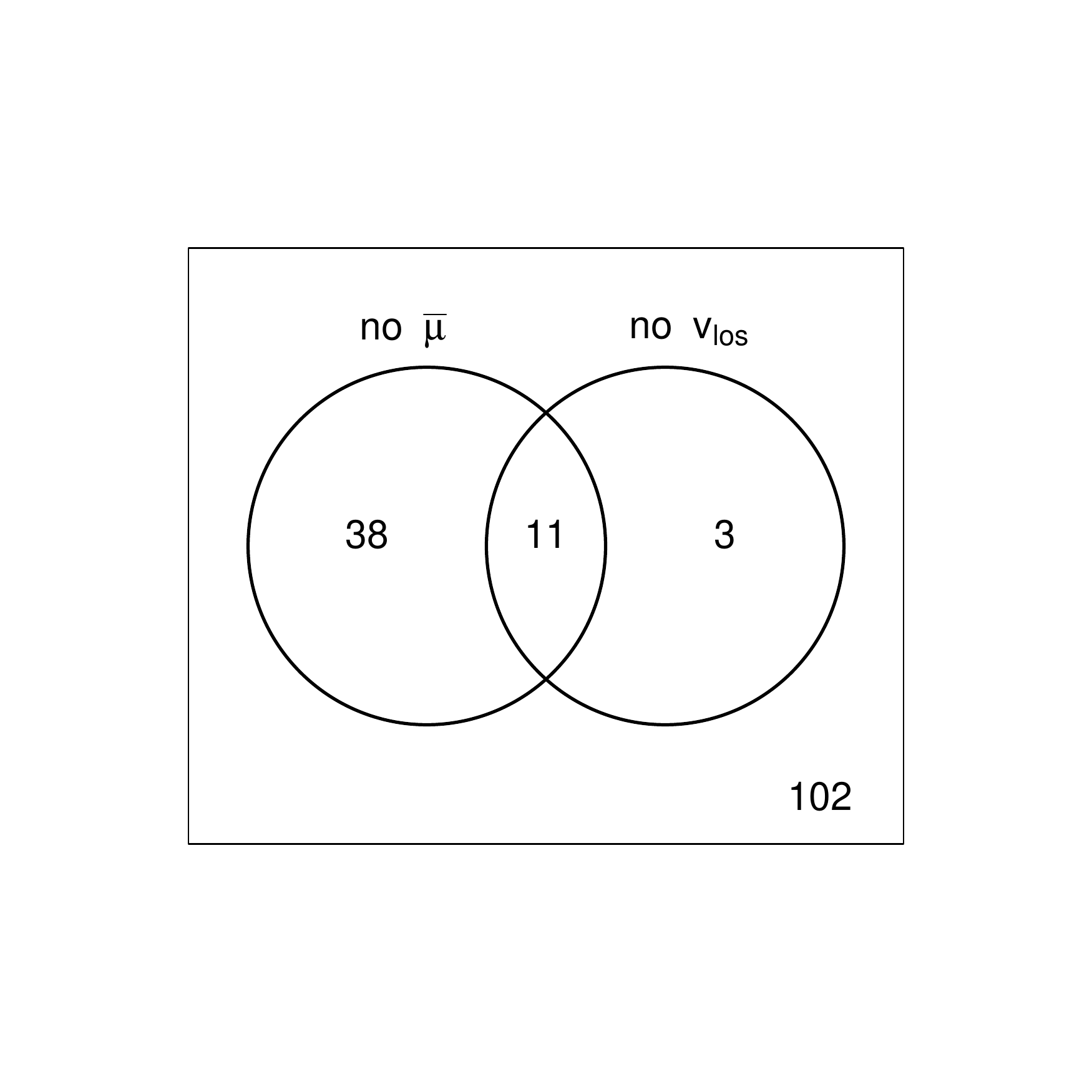}
		\caption{A Venn diagram illustrating the incompleteness of GC data in the compiled Gaia \& HSTPROMO catalog. There are 102 GCs with both line-of-sight velocity ($v_{\text{los}}$) and proper motions ($\bar{\mu}$) measurements (i.e. complete data), 49 GCs without proper motion measurements, and 14 GCs without line-of-sight velocity measurements.}\label{fig:venn}
	\end{figure}

	The total number of GCs in our compiled data set is 157, but four GCs --- Arp 2, Pal 12, Terzan 7, and Terzan 8 --- were recently confirmed by \cite{2018SohnApJ} to be associated with the Sagittarius (Sgr) dwarf galaxy. Thus, \cite{2018SohnApJ} used only one of these GCs (Arp 2) in their study of the MW's mass. They also report no significant change in the mass estimate if Pal 12, Terzan 7, or Terzan 8 is used instead. We follow their lead and exclude all but Arp 2 from our analysis. \cite{2010lawmajewski} listed a few other GCs that may be associated with Sgr, but these have yet to be confirmed and so we leave these GCs in our analysis. Thus, our sample consists of 154 GCs, with a breakdown of complete and incomplete data shown in Figure~\ref{fig:venn}. 
	
	\begin{figure}
	\centering
	\includegraphics[scale=0.4, trim=3cm 3cm 3cm 4cm]{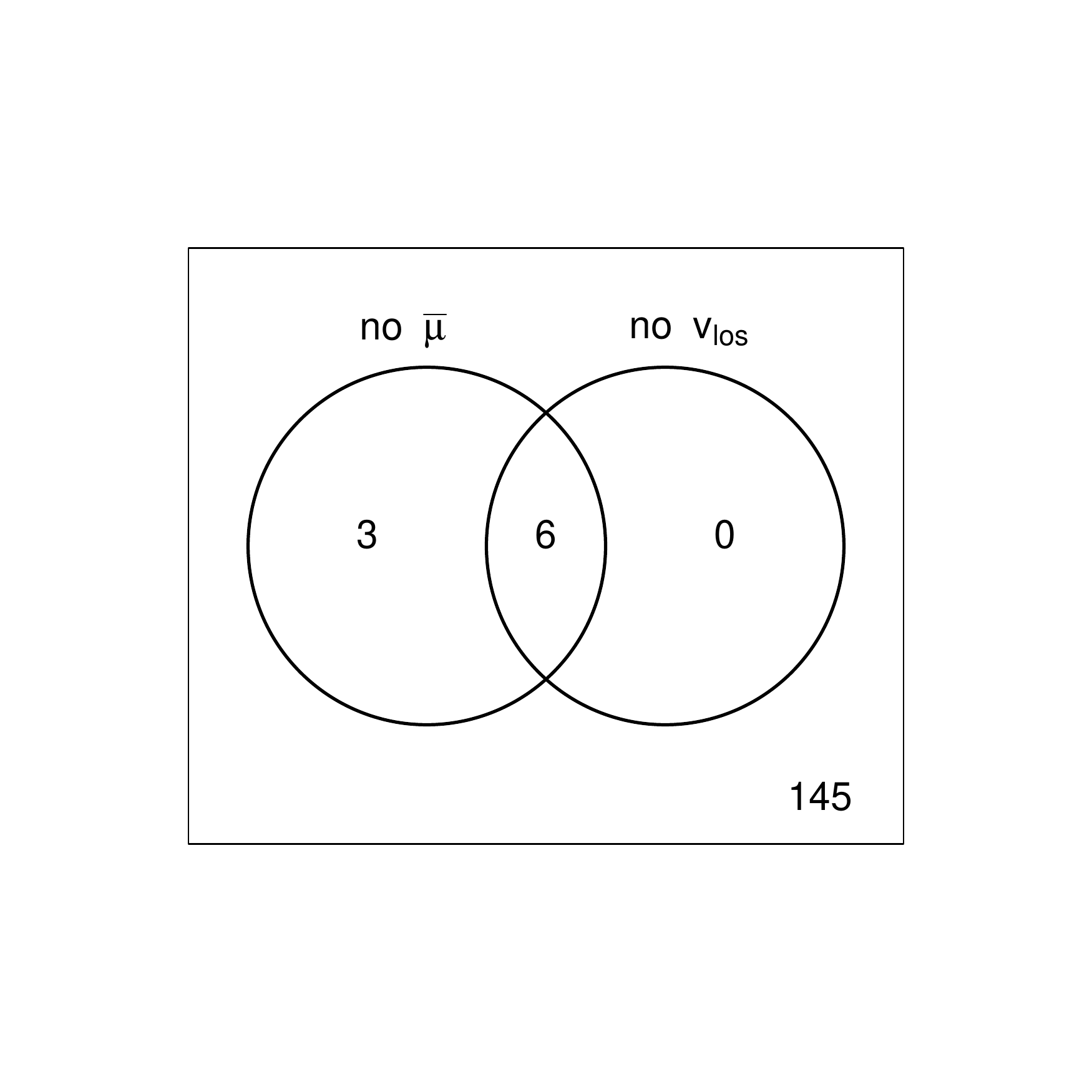}
	\caption{The same as Figure~\ref{fig:venn}, but for the Vasiliev catalog. In this data set there are 145 GCs with complete data, and \textbf{9} GCs without proper motion measurements --- 6 of which also lack line-of-sight velocity measurements.}\label{fig:vennVas}
\end{figure}

	Together, the Gaia collaboration and the HSTPROMO team have greatly increased the number of proper motion measurements of GCs in the MW. Only 52 out of 154 GCs in our updated Gaia $+$ HSTPROMO catalog have incomplete measurements. This is a significant improvement from the data set used previously (Papers 2--3), in which approximately 50\% of the data were incomplete.

	Even more recently, \cite{Vasiliev2019MNRAS} used the Gaia DR2 data to calculate the mean proper motions of 150 GCs in the MW and substantially increase the percentage of complete data. The new Vasiliev catalog is an appealing data set not only because of its completeness, but also because the proper motions are calculated in a consistent manner. Thus, we use this new data set to form a second catalog of GCs and estimate the MW mass and CMP for comparison.
	
	We generate the second catalog from the Vasiliev data such that it contains the same GCs as our Gaia$+$HSTPROMO catalog. We again start with the Harris catalog and then replace all GC measurements of proper motion and line-of-sight velocity with the new measurements reported by Vasiliev \cite[the line-of-sight velocities originate from][]{Baumgardt2018MNRAS}. 
	
	The Vasiliev catalog also contains information for the GC \emph{Leavens 1 (Crater)} and recently discovered GC FSR 1716 \citep{Minniti2017ApJ, Koch2017AA}, but we exclude these data because they are not in the Gaia$+$HSTPROMO catalog. Crater is also the most distant GC known to date \citep{2014leavensCRATER}, and its proper motion measurement is quite uncertain \citep{Vasiliev2019MNRAS}. Although we do not include Crater in this study, its distance could make it an important GC for future MW mass estimates once its proper motion is known with more certainty.
	
	Overall, the main difference between the Gaia $+$ HSTPROMO and Vasiliev catalogs is completeness, with the latter containing complete measurements for 143 GCs, and incomplete measurements for 9 GCs (Figure~\ref{fig:vennVas}).
	
	As mentioned in Section~\ref{sec:methods}, we use GCs with Galactocentric distances $r>15\kpc$, thus limiting our sample size to 35 in both our Gaia $+$ HSTPROMO and Vasiliev catalogs. In this reduced Gaia $+$ HSTPROMO catalog, 16 GCs are missing $\bar{\mu}$ measurements, three of which are also missing $v_{\text{los}}$ measurements. In contrast, the latter three GCs (AM 4, Ko 1, and Ko 2)  are the only incomplete data in the Vasiliev catalog. Because only the distances and positions of AM 4, Ko 1, and Ko 2 are known, these GCs are removed from both catalogs and used to define the $\alpha$ prior distribution's hyperparamters (see Section~\ref{sec:methods}, Figure~\ref{fig:prioralpha}, and Papers 2-3).
	
\begin{figure}
		\centering
		\includegraphics[scale=0.85, trim=1.5cm 1cm 1cm 2cm]{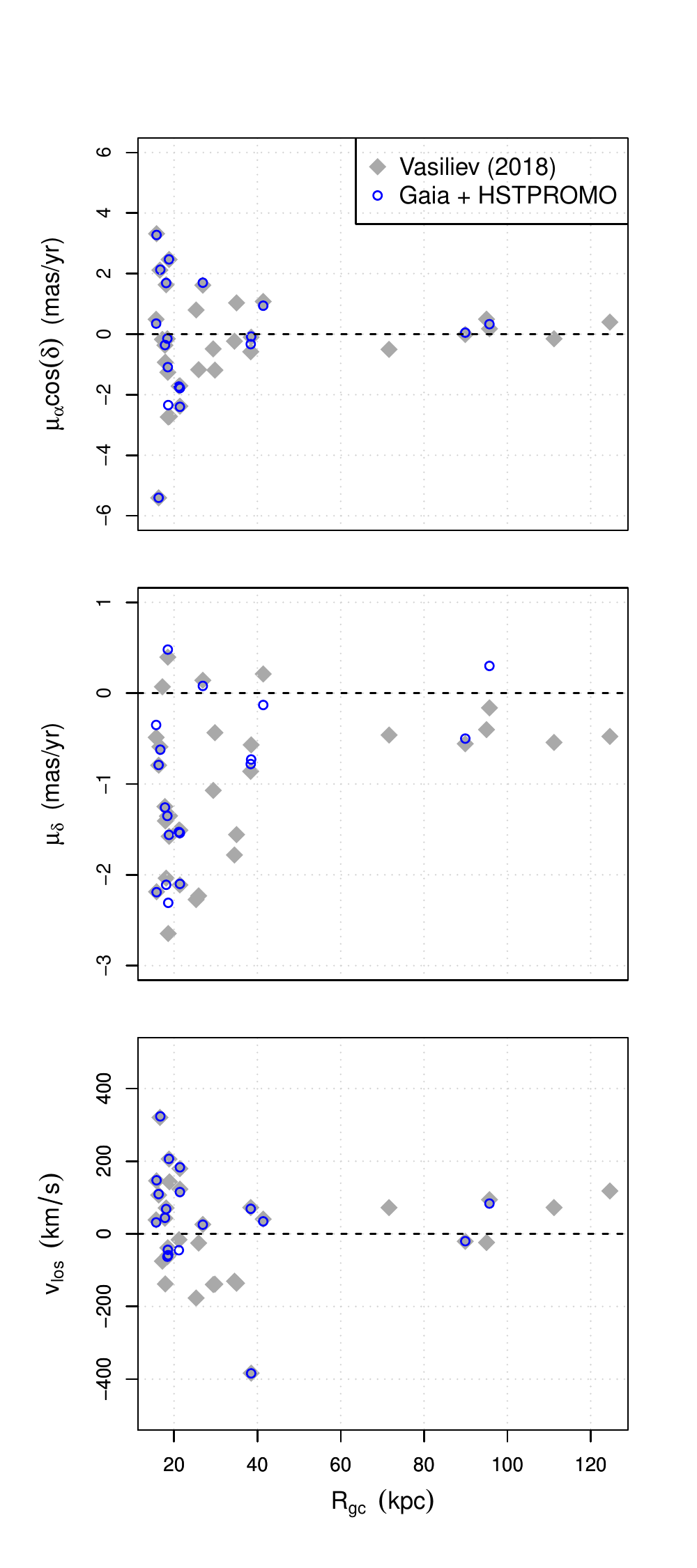}
		\caption{The proper motions ($\mu_{\alpha}\cos\delta$, $\mu_\delta$) and line-of-sight velocities ($v_{\text{los}}$) for GCs beyond 15kpc, as determined by Gaia and HSTPROMO (open blue circless) and by Vasiliev (grey diamonds).}\label{fig:compareGaiaVasiliev}
	\end{figure}

	Figure~\ref{fig:compareGaiaVasiliev} shows the proper motion components and line-of-sight velocities as a function of Galactocentric distance for GCs at $r>15\kpc$. The Gaia$+$HSTPROMO measurements are shown with open blue circles and the Vasiliev's measurements are grey diamonds. There is excellent agreement between the catalogs' measurements. The only outlier is the $\mu_\delta$ measurements for Pal 3, which can be seen in the second panel at 95.7\kpc.
	
	\section{Results \& Discussion}\label{sec:results}

\begin{figure*}
	 	\centering
	 	
	 	\includegraphics[scale=0.6]{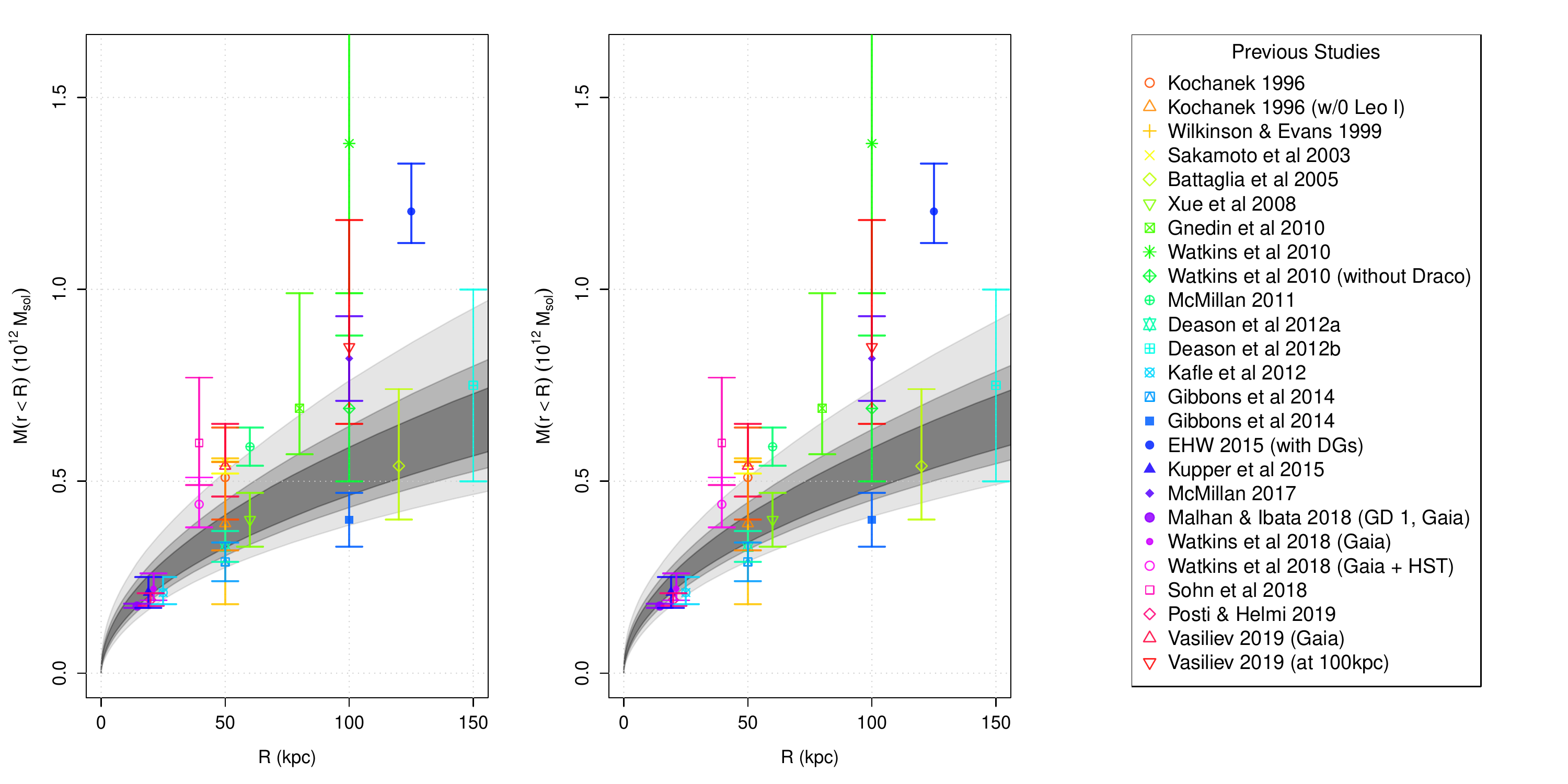}
	 	\caption{The CMPs of the MW with 50\%, 75\%, and 95\% Bayesian credible regions (grey regions), given the Gaia + HSTPROMO data (left), and the Vasiliev data (center). The points with error bars are mass estimates from other studies, which are listed in the legend (right). Marginal distributions for the mass within any $r$ are easily calculated, with some specific examples shown in Figure~\ref{fig:marginals}.}
	 	\label{fig:CMPs}
	 	
	 \end{figure*}

	 Figure~\ref{fig:CMPs} shows the CMP of the MW with Bayesian credible regions (50\%, 75\%, and 95\%) given the Gaia + HSTPROMO catalog (left) and the Vasiliev catalog (center). These CMPs were calculated using the posterior distribution of model parameters (Section~\ref{sec:CMP}), and GCs at $r>15\kpc$.  The points with error bars in Figure~\ref{fig:CMPs} are mass estimates reported in other studies, which are identified in the legend \citep{kochanek1996, wilkinsonevans1999, sakamoto2003, battaglia2005, xue2008, gnedin2010, watkins2010, mcmillan2011, deason2012, deasonetal2012MNRAS, kafle2012, gibbons2014, 2015EHW, kupper2015}, including five which also use Gaia DR2 data \citep{malhanibata2018arXiv, 2019PostiHelmiAA, sohn2018IAUS, Watkins2019ApJ, Vasiliev2019MNRAS}.

	 The two CMPs in Figure~\ref{fig:CMPs} are nearly identical, but the Vasiliev catalog provides a slightly better constraint on the mass because the data are more complete. The median estimates of $M_{200}$ given each catalog are also extremely similar at
	 \vspace{-1ex}
	 \begin{equation}
		0.70~(0.47, 1.14) \times 10^{12} \msun ~~ \text{(Gaia + HSTPROMO)}
	\end{equation}
		and
		\begin{equation}	\label{eq:M200}
		0.70~(0.51, 1.10) \times 10^{12} \msun ~~ \text{~(Vasiliev),}
	 \end{equation}	 
	where numbers in brackets are the lower and upper bounds of the 95\% Bayesian credible region. The quantiles for the model parameters calculated from the posterior distributions are also extremely similar between the two data sets. 	 Henceforth, we refer to the Vasiliev  results in the discussion. 
	 
	A large number of GCs in our catalogs lie between 15 and 20kpc of the Galactic centre (Figure~\ref{fig:compareGaiaVasiliev}), and yet we extrapolate the results out to larger distances with the CMP. To test how sensitive our method is to these inner GCs, we perform the analysis again on the Vasiliev catalog, using a new cut of $r>20\kpc$.

\begin{figure}
	 	\centering
	 	\includegraphics[scale=0.5]{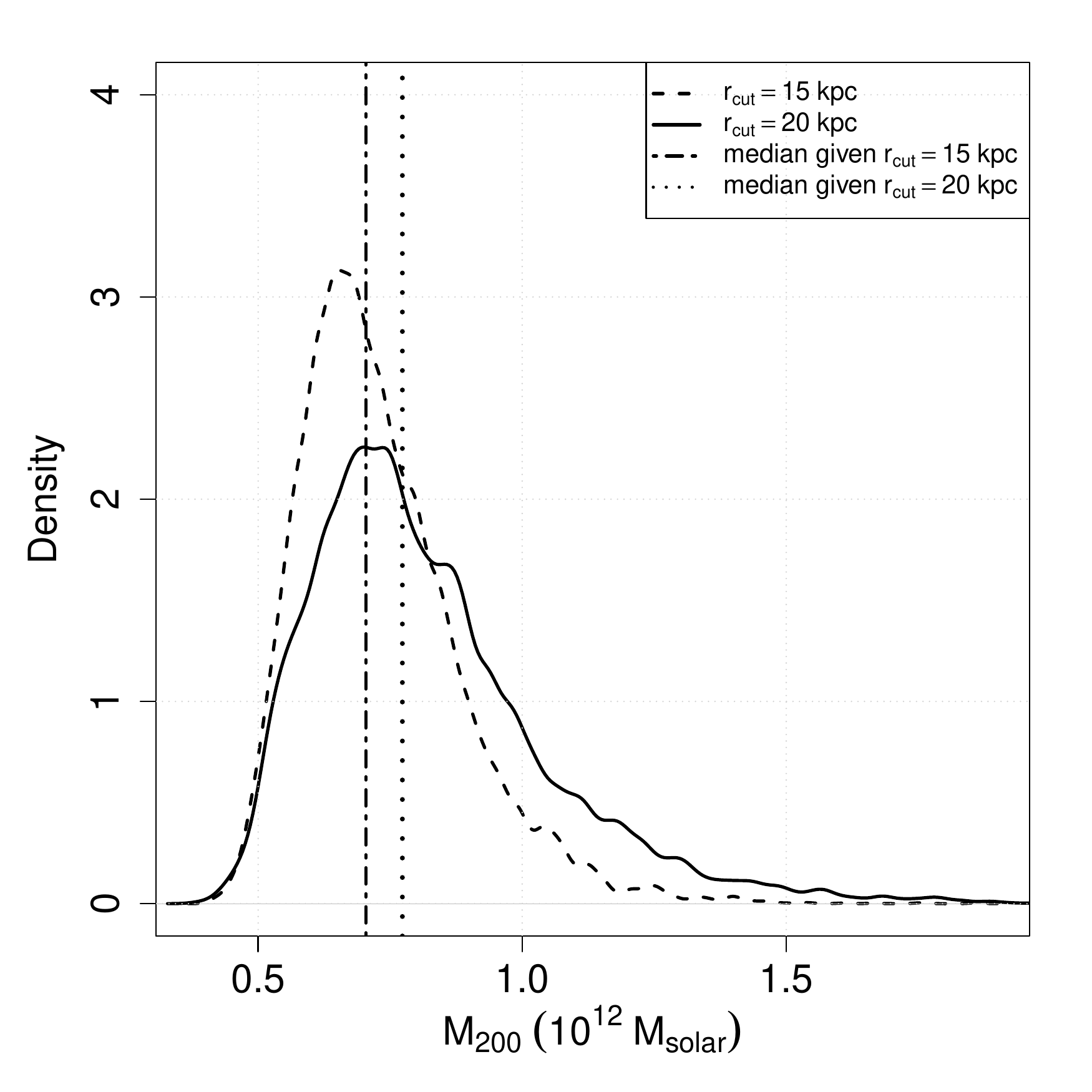}
	 	\caption{The marginal posterior distributions (curves) and medians (vertical lines) for $M_{200}$ of the MW, given the Vasiliev data. The two distributions and their medians represent the results when data at $r>15\kpc$ and $r>20\kpc$ are used. The median values are shown in the first and fourth row of Table~\ref{tab:M200}.}\label{fig:M200marg}
	 	
	 \end{figure}

	Figure~\ref{fig:M200marg} compares the marginal posterior distributions and medians for $M_{200}$, given $r_{\text{cut}}$ values of 15kpc and 20\kpc. The marginal distributions are very similar, but the median estimate for $M_{200}$ and the upper bound of the 95\% credible interval are slightly higher: 
	\begin{equation}
		0.77~ (0.51,1.40)\times 10^{12}\msun \text{~~($r_{\text{cut}}$=20\kpc)}.
	\end{equation}
The new CMP is virtually unchanged barring the larger upper values for the Bayesian credible regions, indicating that the method is relatively insensitive to the tracers between 15kpc and 20kpc.
	 
Recently, \cite{Myeongetal2018ApJL} found evidence that up to ten GCs (NGCs 362, 1261, 1851, 1904, 2298, 2808, 5286, 6779, 6864, and 7089) were brought into the Galaxy during a massive merger event \citep{Belokurov2018MNRAS, Kruijssen2018}. All of these GCs reside within 20kpc of the Galaxy, and thus the sensitivity test just mentioned has excluded them (Figure~\ref{fig:M200marg}). Four of Myeong's GCs, however, do reside between 15 and 20\kpc, so testing their possible influence on the mass estimate when $r_{\text{cut}}=15\kpc$ is a worthwhile exercise.
	 
	 
\begin{table*}[ht]
\centering
\begin{tabular}{cccc}
  \hline
  Data used  & & $M_{200}$ ($10^{12}\msun$) & \\
  \hline
  \hline
 & Lower 95\% c.r. & Median & Upper 95\% c.r. \\ 
  \hline
$r>15$ kpc & 0.51 & 0.70 & 1.10 \\ 
  $r>15$ kpc, Myeong GCs removed & 0.53 & 0.77 & 1.23 \\ 
  $r>15$ kpc, 4 random GCs removed & 0.49 & 0.70 & 1.11 \\ 
  $r>20$ kpc & 0.51 & 0.77 & 1.40 \\ 
   \hline
\end{tabular}
\caption{Mass estimates and 95\% credible regions for $M_{200}$ ($10^{12}\msun$) given different subsets of the Vasiliev data.} 
\label{tab:M200}
\end{table*}

	 Interestingly, when the four Myeong GCs between $15\kpc<r<20\kpc$ are excluded from the analysis, the median estimate of $M_{200}$ increases slightly from Eq.~\ref{eq:M200} to
	 \begin{equation}
	 	0.77~ (0.53,1.23)\times 10^{12}\msun. 
	 \end{equation}
	Yet, when we put these GCs back into the analysis and remove four other GCs between $15\kpc < r < 20\kpc$ at random, the median estimate is,
	\begin{equation}
		0.70~(0.49, 1.11)\times 10^{12}\msun. 
	\end{equation}
	This result seems to suggest that the Myeong's GCs push our mass estimate to smaller values. Although this is intriguing, the effect is minor when we consider the width of the Bayesian credible regions. We summarize the results of this sensitivity test in Table~\ref{tab:M200}.

\begin{figure}
	 	\centering
	 	\includegraphics[scale=0.8, trim=1cm 0cm 0cm 0cm]{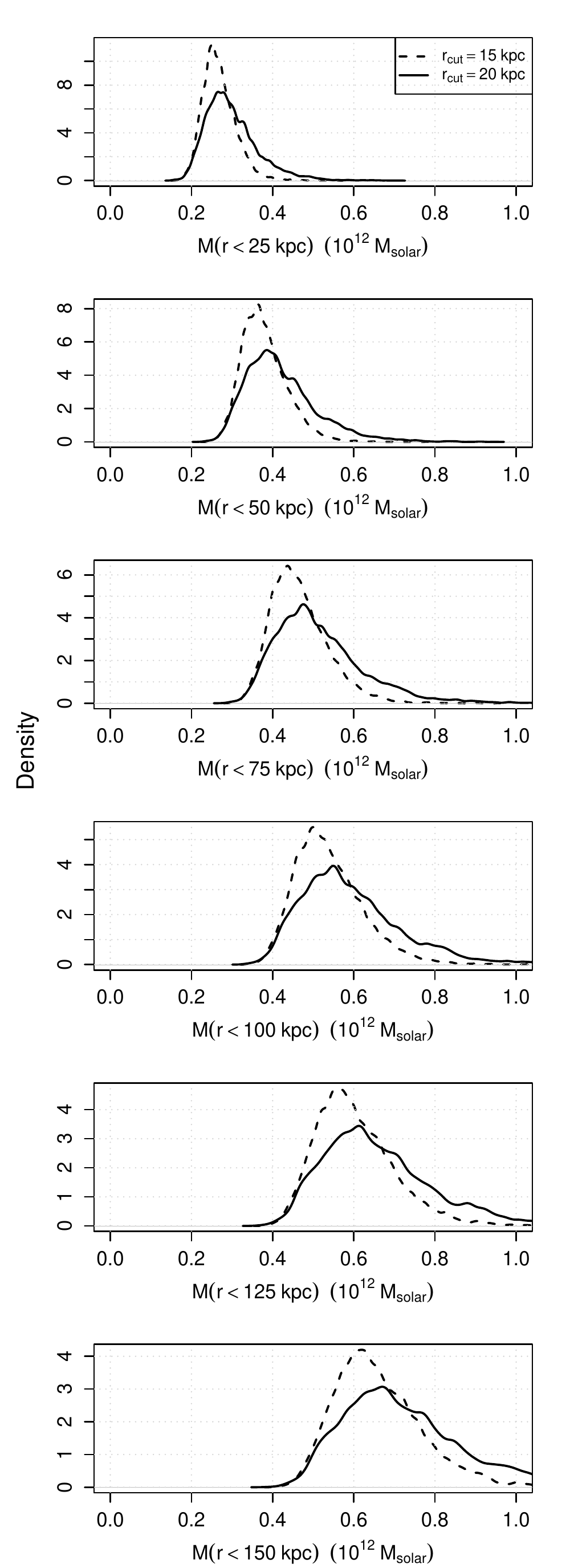}
	 	\caption{Marginal distributions for the mass within different radii, from $r<25\kpc$ (top) to $r<150\kpc$ (bottom) in steps of $25\kpc$. The two curves shown in each plot correspond to different $r_{\text{cut}}$ values. }\label{fig:marginals}
	 \end{figure}

	The main advantage of our CMP for the MW is the ability to easily compare our mass estimate to many other studies both visually (i.e. Figure~\ref{fig:CMPs}) and quantitatively, within any distance from the Galactic center. To illustrate the quantitative comparisons, we present slices of our CMP $M(<r)$ taken at $r=25\kpc$ through $r=150\kpc$, by steps of 25kpc, to show the marginal posterior distributions for $(M<r)$ at those distances (Figure~\ref{fig:marginals}). Marginal distributions for the mass within any $r$ are also easily calculated. 
	 
	Our CMP agrees with many of the results reported in the literature within their uncertainties, except for three higher-mass estimates in Figure~\ref{fig:CMPs}: \cite{watkins2010} (with Draco) at 100kpc,  \cite{2015EHW} at 125kpc, and \cite{Vasiliev2019MNRAS} at 100kpc. Notably, the first two of these studies included dwarf galaxies in their samples, whereas we do not. There is evidence suggesting that some dwarf galaxies may inflate the mass estimate of the MW. Case in point, the result of \cite{watkins2010} without Draco is in good agreement with our estimate at 100kpc.
 
	Other studies that use dwarf galaxies, or even high/extreme velocity stars, also tend to infer ``heavier'' mass estimates of the MW. Many of these studies' results cannot be included on the CMP (Figure~\ref{fig:CMPs}) because they are reported as $M_{200}$ or the virial mass $M_{\text{vir}}$. Instead, we compare three of these studies' results to our marginal distribution for $M_{200}$ given the Vasiliev catalog (Figure~\ref{fig:M200marg}).
	 
	\cite{patel2017MNRAS} performed a Bayesian analysis in conjunction with cosmological simulations of MW-type galaxies to infer the Galaxy's mass. They used the kinematics of one massive satellite, and inferred $M_{200}=1.223\times10^{12}\msun$ and a 95\% Bayesian credible of $(1.207, 1.248)\times10^{12}\msun$. \cite{2018hattoriApJ} also deduced a high mass using newly discovered extreme velocity stars in the Gara DR2 data. If these stars are bound to our Galaxy, then they imply $M_{200}\approx 1.4\times10^{12}$\msun. As a final example, \cite{monari2018AA} used counter-rotating halo stars in Gaia to estimate the escape velocity of the MW between 5-10kpc. Their results also implied a heavy MW with a dark matter virial mass of $M_{200} = 1.55^{+0.64}_{-0.51} \times 10^{12}\msun$. All three of these mass estimates are almost completely ruled out by our estimate of $M_{200}$ (Figure~\ref{fig:M200marg}), which did not include high/extreme velocity stars or dwarf galaxies.

	It is less clear why the estimate at 100kpc by \cite{Vasiliev2019MNRAS} is slightly higher than our estimate, although the uncertainties do overlap with our 95\% c.r.. Two major differences exist between our studies. The first difference is that they used all GCs in their catalog, whereas we only used GCs beyond 15kpc. The second difference is the choice of gravitational potential model. While we chose a simple power-law model and excluded inner GCs in our analysis, they used a three-component model and included GCs at all distances.
	
	We also compare our results to studies that use GC data from Gaia and/or HSTPROMO. For example, \cite{2018SohnApJ} estimated the MW mass within 39.5kpc using 16 GCs with proper motions measurements by HSTPROMO. Utilizing the estimator from \cite{watkins2010} and Monte Carlo simulations to estimate the uncertainty, they find $M(r<39.5\kpc) = 0.61^{+0.18}_{-0.12}\times 10^{12}\msun$. Our estimate for the mass within 39.5kpc is
	\begin{equation}
		M(r<39.5\kpc) = 0.33~(0.26, 0.45)\times 10^{12} \msun,
		\label{eq:M39}
	\end{equation} 
	which does not agree with their results within the uncertainties. However, more recently \cite{Watkins2019ApJ} combined Gaia DR2 data with HSTPROMO data, and used the same estimator as \cite{2018SohnApJ}. They found $M(r<39.5\kpc)=0.44^{0.07}_{-0.06}\times 10 ^{12}\msun$. Reassuringly, the latter agrees with our estimate at this distance within the uncertainties, despite their use of a different model for $M(<r)$. 
	
	\cite{2019PostiHelmiAA} also use the kinematic data of 75 GCs provided by Gaia DR2 and HSTPROMO to estimate the mass of the MW in a Bayesian analysis. In contrast to this study, they use a two-component DF for the GC system of the MW. The assumption is that the GC population dynamics can be decomposed into disk-like and halo-like components, with the latter containing parameters for the mean rotational velocity and velocity anisotropy of the GC system. They also include a shape parameter for the dark matter halo. Because their study uses a Bayesian paradigm, like ours, their method includes measurements uncertainties and incomplete data directly in the analysis. Despite using a more complicated DF than this study, their mass estimate within 20kpc
	($M(r< 20 \kpc) = 0.191^{+0.17}_{-0.15} \times 10^{12} \msun$) is within the bounds of our 95\% Bayesian credible regions:
	\begin{equation*}
		M(r<20\kpc) = 0.24~(0.18,0.32) \times 10^{12} \msun. 
	\end{equation*}

	\cite{Vasiliev2019MNRAS} estimates the mass of the MW using his GC catalog in a maximum likelihood analysis. Vasiliev's analysis assumes a single-component DF for the GC population, but a three-component gravitational potential to account for the bulge, disc, and halo components. He finds $M(r<50\kpc)= 0.54^{+0.11}_{-0.08}\times 10^{12}\msun$. The uncertainties overlap with the upper bound of the 95\% Bayesian credible for our mass estimate within 50kpc:
	\begin{equation*}
		M(r<50\kpc) = 0.37~(0.29, 0.51)\times 10^{12}\msun.
	\end{equation*}

Finally, \cite{malhanibata2018arXiv} used Gaia data for the GD-1 stellar stream and found $M(r< 14.5\kpc) = 0.175^{+0.06}_{-0.05} \times 10^{12}\msun$, which is in agreement with
\begin{equation}
	M(r< 14.5\kpc) = 0.20~(0.15, 0.28) \times 10^{12}\msun
\end{equation}
given by our method.

Overall, it appears that studies are reaching a consensus on the mass of the MW within 40kpc since the release of Gaia DR2 and HSTPROMO. Even when different analysis methods and models are confronted with the same data, similar conclusions are reached. Albeit, beyond 15kpc the broken power-law models used in \cite{Vasiliev2019MNRAS} and \cite{2019PostiHelmiAA} are similar to our single power-law model, so the agreement here is perhaps less surprising.
	
Mass estimates of the Galaxy out to larger distances, however, are still rather uncertain. For example, \cite{2018SohnApJ} and \cite{Watkins2019ApJ} report virial mass estimates of $M_{\text{vir}} = 2.05^{+0.97}_{-0.79}\times10^{12}\msun$ and $M_{\text{vir}} = 1.41^{+1.99}_{-0.52}\times10^{12}\msun$ respectively. 	\cite{2019PostiHelmiAA} provide a tighter constraint on the virial mass at $M_{\text{vir}} = 1.3\pm 0.3 \times 10^{12} \msun$, as does our Bayesian estimate of $M_{200}$ (Equation~\ref{eq:M200}). Although there is some tension between results, our marginal credible regions for $M_{200}$ (Figure~\ref{fig:M200marg}) do overlap with those of \cite{2019PostiHelmiAA}.

\section{Conclusions \& Future Work}\label{sec:conclude}

The state of knowledge about the MW's total mass is still uncertain, but the state of data for tracers has vastly improved. Multiple studies have now estimated the mass within 40kpc, and in general they are in good agreement with one another. Our CMP of the MW within $r<50\kpc$ is also in agreement with many recent results. 

Our CMP overlaps more than half of the mass estimates at different radii, suggesting a less massive MW. However, mass estimates that disagree with our results tend to be at larger distances. These discrepancies may be attributed to the inclusion/exclusion of tracers such as dwarf galaxies and/or high velocity stars in the analyses, but it could also be due to differing model assumptions. Thus, the Galaxy's mass within larger distances is still up for debate. Until more complete measurements for tracers at larger distances are acquired, the total mass of the MW will remain uncertain. 

Comparing and verifying mass estimates at large distances will remain challenging without more data, because of the variety of models and data analysis techniques. As a community, we compare results by looking at the uncertainties presented in each others' studies, with the caveat that everyone's methods and data are different. We have seen how the Gaia DR2 and HSTPROMO data have greatly improved our understanding of the mass within 40kpc of the MW within this context. Thus, it is imperative to obtain complete velocity measurements of tracers at large distances to reach a similar consensus about the Galaxy's total mass.

At the same time, studies continue to report mass estimates within different distances from the Galactic center, making comparisons difficult. The CMP presented in this work instead provides a holistic view of our understanding and estimate of the MW's mass at all distances. Thus, we hope CMPs with Bayesian credible regions become a standard way to present and compare estimates of the MW's mass in future studies, and openly share instructions to create these profiles.

In summary, future work on the mass of the MW will benefit greatly from complete velocity measurements of tracers at greater distances, and by comparing CMPs from different studies.

	\acknowledgments
	
	The authors thank the anonymous referee for their insightful and helpful report.  G.~Eadie would like to extend gratitude to E.~Vasiliev for sharing his catalog of new proper motions measurements and for useful discussions.
	
	This research was funded by an eScience Institute Postdoctoral Fellowship to G.~Eadie, made possible from generous donations from the three Moore, Sloan, and Washington Research Foundations at the eScience Institute, University of Washington (UW). G.~Eadie and M.~Juric acknowledge support from the University of Washington Institute for Data Intensive Research in Astrophysics and Cosmology (DIRAC). The DIRAC Institute is supported through generous gifts from the Charles and Lisa Simonyi Fund for Arts and Sciences, and the Washington Research Foundation. M.~Juric acknowledges the support of the Washington Research Foundation Data Science Team Chair fund, and the UW Provost's Initiative in Data-Intensive Discovery. The authors would like to thank members of the DIRAC Institute for insightful conversations regarding this work.  
	
	This work uses the data from the European Space Agency mission Gaia (\url{https://www.cosmos.esa.int/gaia}), processed by the Gaia Data Processing and Analysis Consortium (\url{https://www.cosmos.esa.int/web/gaia/dpac/consortium}). 

	\software{R Statistical Software Environment \citep{R}, including the following packages:  \emph{stats} (part of base R), \emph{CODA:  Convergence Diagnosis and Output Analysis for MCMC} \citep{coda, codapackage},   \emph{emdbook: Ecological Models and Data in R} (\citealt{emdbookBOOK}, \citeyear{emdbook} version), \emph{ggplot2} \citep{hadleyGGPLOT}, \emph{limma} (vennDiagram function)  \citep{limmapackage}, \emph{MASS: Modern Applied Statistics with S} \cite{mass}, \emph{moments: Moments, cumulants, skewness, kurtosis and related tests} \citep{moments}, \emph{pracma: Practical Numerical Math Functions} \citep{pracma}, \emph{RColorBrewer: ColorBrewer Palettes}, and \emph{SNOW: Simple Network of Workstations} \citep{snow}. \citep{rcolorbrewer}. }

	\bibliographystyle{aasjournal}
	\bibliography{refs}

\begin{thebibliography}{}
\expandafter\ifx\csname natexlab\endcsname\relax\def\natexlab#1{#1}\fi
\providecommand{\url}[1]{\href{#1}{#1}}
\providecommand{\dodoi}[1]{doi:~\href{http://doi.org/#1}{\nolinkurl{#1}}}
\providecommand{\doeprint}[1]{\href{http://ascl.net/#1}{\nolinkurl{http://ascl.net/#1}}}
\providecommand{\doarXiv}[1]{\href{https://arxiv.org/abs/#1}{\nolinkurl{https://arxiv.org/abs/#1}}}

\bibitem[{{Battaglia} {et~al.}(2005){Battaglia}, {Helmi}, {Morrison},
  {Harding}, {Olszewski}, {Mateo}, {Freeman}, {Norris}, \&
  {Shectman}}]{battaglia2005}
{Battaglia}, G., {Helmi}, A., {Morrison}, H., {et~al.} 2005, \mnras, 364, 433,
  \dodoi{10.1111/j.1365-2966.2005.09367.x}

\bibitem[{{Baumgardt} \& {Hilker}(2018)}]{Baumgardt2018MNRAS}
{Baumgardt}, H., \& {Hilker}, M. 2018, \mnras, 478, 1520,
  \dodoi{10.1093/mnras/sty1057}

\bibitem[{{Baumgardt} {et~al.}(2019){Baumgardt}, {Hilker}, {Sollima}, \&
  {Bellini}}]{Baumgardt2019MNRAS}
{Baumgardt}, H., {Hilker}, M., {Sollima}, A., \& {Bellini}, A. 2019, \mnras,
  482, 5138, \dodoi{10.1093/mnras/sty2997}

\bibitem[{{Belokurov} {et~al.}(2018){Belokurov}, {Erkal}, {Evans}, {Koposov},
  \& {Deason}}]{Belokurov2018MNRAS}
{Belokurov}, V., {Erkal}, D., {Evans}, N.~W., {Koposov}, S.~E., \& {Deason},
  A.~J. 2018, \mnras, 478, 611, \dodoi{10.1093/mnras/sty982}

\bibitem[{Bolker(2018)}]{emdbook}
Bolker, B. 2018, emdbook: Ecological Models and Data in {R}

\bibitem[{Bolker(2008)}]{emdbookBOOK}
Bolker, B.~M. 2008, Ecological Models and Data in R (Princeton University
  Press)

\bibitem[{Borchers(2017)}]{pracma}
Borchers, H.~W. 2017, pracma: Practical Numerical Math Functions

\bibitem[{{Casetti-Dinescu} {et~al.}(2013){Casetti-Dinescu}, {Girard},
  {J{\'{\i}}lkov{\'a}}, {van Altena}, {Podest{\'a}}, \&
  {L{\'o}pez}}]{dinescu2013}
{Casetti-Dinescu}, D.~I., {Girard}, T.~M., {J{\'{\i}}lkov{\'a}}, L., {et~al.}
  2013, \aj, 146, 33, \dodoi{10.1088/0004-6256/146/2/33}

\bibitem[{{Casetti-Dinescu} {et~al.}(2010){Casetti-Dinescu}, {Girard},
  {Korchagin}, {van Altena}, \& {L{\'o}pez}}]{dinescu2010}
{Casetti-Dinescu}, D.~I., {Girard}, T.~M., {Korchagin}, V.~I., {van Altena},
  W.~F., \& {L{\'o}pez}, C.~E. 2010, \aj, 140, 1282,
  \dodoi{10.1088/0004-6256/140/5/1282}

\bibitem[{{de la Fuente Marcos} {et~al.}(2015){de la Fuente Marcos}, {de la
  Fuente Marcos}, {Moni Bidin}, {Ortolani}, \&
  {Carraro}}]{delaFuenteMarcos2015AA}
{de la Fuente Marcos}, R., {de la Fuente Marcos}, C., {Moni Bidin}, C.,
  {Ortolani}, S., \& {Carraro}, G. 2015, \aap, 581, A13,
  \dodoi{10.1051/0004-6361/201526580}

\bibitem[{{Deason} {et~al.}(2012{\natexlab{a}}){Deason}, {Belokurov}, {Evans},
  \& {An}}]{deason2012}
{Deason}, A.~J., {Belokurov}, V., {Evans}, N.~W., \& {An}, J.
  2012{\natexlab{a}}, \mnras, 424, L44,
  \dodoi{10.1111/j.1745-3933.2012.01283.x}

\bibitem[{{Deason} {et~al.}(2012{\natexlab{b}}){Deason}, {Belokurov}, {Evans},
  \& {McCarthy}}]{Deason2012ApJ}
{Deason}, A.~J., {Belokurov}, V., {Evans}, N.~W., \& {McCarthy}, I.~G.
  2012{\natexlab{b}}, \apj, 748, 2, \dodoi{10.1088/0004-637X/748/1/2}

\bibitem[{{Deason} {et~al.}(2012{\natexlab{c}}){Deason}, {Belokurov}, {Evans},
  {Koposov}, {Cooke}, {Pe{\~n}arrubia}, {Laporte}, {Fellhauer}, {Walker}, \&
  {Olszewski}}]{deasonetal2012MNRAS}
{Deason}, A.~J., {Belokurov}, V., {Evans}, N.~W., {et~al.} 2012{\natexlab{c}},
  \mnras, 425, 2840, \dodoi{10.1111/j.1365-2966.2012.21639.x}

\bibitem[{{Eadie} \& {Harris}(2016)}]{eadie2016}
{Eadie}, G., \& {Harris}, W. 2016, \apj, 829, 108,
  \dodoi{10.3847/0004-637X/829/2/108}

\bibitem[{{Eadie} {et~al.}(2015){Eadie}, {Harris}, \& {Widrow}}]{2015EHW}
{Eadie}, G., {Harris}, W., \& {Widrow}, L. 2015, \apj, 806, 54,
  \dodoi{10.1088/0004-637X/806/1/54}

\bibitem[{{Eadie} {et~al.}(2018){Eadie}, {Keller}, \& {Harris}}]{EKH2018}
{Eadie}, G.~M., {Keller}, B.~W., \& {Harris}, W.~E. 2018, \apj, 865, 72

\bibitem[{{Eadie} {et~al.}(2017){Eadie}, {Springford}, \& {Harris}}]{ESH2017}
{Eadie}, G.~M., {Springford}, A., \& {Harris}, W.~E. 2017, \apj, 835, 167,
  \dodoi{10.3847/1538-4357/835/2/167}

\bibitem[{{Evans} {et~al.}(1997){Evans}, {Hafner}, \& {de Zeeuw}}]{evans1997}
{Evans}, N.~W., {Hafner}, R.~M., \& {de Zeeuw}, P.~T. 1997, \mnras, 286, 315,
  \dodoi{10.1093/mnras/286.2.315}

\bibitem[{Evans {et~al.}(2016)Evans, Sanders, Williams, An, Lynden-Bell, \&
  Dehnen}]{2016WilliamsetalMNRAS}
Evans, N.~W., Sanders, J.~L., Williams, A.~A., {et~al.} 2016, \mnras, 456,
  4506, \dodoi{10.1093/mnras/stv2729}

\bibitem[{{Feltzing} \& {Johnson}(2002)}]{feltzing2002}
{Feltzing}, S., \& {Johnson}, R.~A. 2002, \aap, 385, 67,
  \dodoi{10.1051/0004-6361:20011771}

\bibitem[{{Fritz} \& {Kallivayalil}(2015)}]{fritz2015}
{Fritz}, T.~K., \& {Kallivayalil}, N. 2015, \apj, 811, 123,
  \dodoi{10.1088/0004-637X/811/2/123}

\bibitem[{{Gaia Collaboration} {et~al.}(2018){Gaia Collaboration}, {Helmi},
  {van Leeuwen}, {McMillan}, {Massari}, {Antoja}, {Robin}, {Lindegren},
  {Bastian}, {Arenou}, \& et~al.}]{2018GaiaDR2HelmiGCS}
{Gaia Collaboration}, {Helmi}, A., {van Leeuwen}, F., {et~al.} 2018, \aap, 616,
  A12, \dodoi{10.1051/0004-6361/201832698}

\bibitem[{{Gibbons} {et~al.}(2014){Gibbons}, {Belokurov}, \&
  {Evans}}]{gibbons2014}
{Gibbons}, S.~L.~J., {Belokurov}, V., \& {Evans}, N.~W. 2014, \mnras, 445,
  3788, \dodoi{10.1093/mnras/stu1986}

\bibitem[{{Gnedin} {et~al.}(2010){Gnedin}, {Brown}, {Geller}, \&
  {Kenyon}}]{gnedin2010}
{Gnedin}, O.~Y., {Brown}, W.~R., {Geller}, M.~J., \& {Kenyon}, S.~J. 2010,
  \apjl, 720, L108, \dodoi{10.1088/2041-8205/720/1/L108}

\bibitem[{{Harris}(1996)}]{1996harrisPaper}
{Harris}, W.~E. 1996, \aj, 112, 1487, \dodoi{10.1086/118116}

\bibitem[{{Harris}(2010)}]{2010Harris}
---. 2010, astro-ph.
\newblock \doarXiv{1012.3224}

\bibitem[{Hattori {et~al.}(2018)Hattori, Valluri, Bell, \&
  Roederer}]{2018hattoriApJ}
Hattori, K., Valluri, M., Bell, E.~F., \& Roederer, I.~U. 2018, \apj, 866, 121,
  \dodoi{10.3847/1538-4357/aadee5}

\bibitem[{{Kafle} {et~al.}(2012){Kafle}, {Sharma}, {Lewis}, \&
  {Bland-Hawthorn}}]{kafle2012}
{Kafle}, P.~R., {Sharma}, S., {Lewis}, G.~F., \& {Bland-Hawthorn}, J. 2012,
  \apj, 761, 98, \dodoi{10.1088/0004-637X/761/2/98}

\bibitem[{{Koch} {et~al.}(2017){Koch}, {Kunder}, \& {Wojno}}]{Koch2017AA}
{Koch}, A., {Kunder}, A., \& {Wojno}, J. 2017, \aap, 605, A128,
  \dodoi{10.1051/0004-6361/201731771}

\bibitem[{Kochanek(1996)}]{kochanek1996}
Kochanek, C.~S. 1996, \apj, 457, 228

\bibitem[{Komsta \& Novomestky(2015)}]{moments}
Komsta, L., \& Novomestky, F. 2015, moments: Moments, cumulants, skewness,
  kurtosis and related tests

\bibitem[{{Kruijssen} {et~al.}(2018){Kruijssen}, {Pfeffer}, {Reina-Campos},
  {Crain}, \& {Bastian}}]{Kruijssen2018}
{Kruijssen}, J.~M.~D., {Pfeffer}, J.~L., {Reina-Campos}, M., {Crain}, R.~A., \&
  {Bastian}, N. 2018, \mnras, 1537, \dodoi{10.1093/mnras/sty1609}

\bibitem[{{K{\"u}pper} {et~al.}(2015){K{\"u}pper}, {Balbinot}, {Bonaca},
  {Johnston}, {Hogg}, {Kroupa}, \& {Santiago}}]{kupper2015}
{K{\"u}pper}, A.~H.~W., {Balbinot}, E., {Bonaca}, A., {et~al.} 2015, \apj, 803,
  80, \dodoi{10.1088/0004-637X/803/2/80}

\bibitem[{Laevens {et~al.}(2014)Laevens, Martin, Sesar, Bernard, Rix, Slater,
  Bell, Ferguson, Schlafly, Burgett, Chambers, Denneau, Draper, Kaiser,
  Kudritzki, Magnier, Metcalfe, Morgan, Price, Sweeney, Tonry, Wainscoat, \&
  Waters}]{2014leavensCRATER}
Laevens, B. P.~M., Martin, N.~F., Sesar, B., {et~al.} 2014, \apjl, 786, L3,
  \dodoi{10.1088/2041-8205/786/1/l3}

\bibitem[{{Law} \& {Majewski}(2010)}]{2010lawmajewski}
{Law}, D.~R., \& {Majewski}, S.~R. 2010, \apj, 718, 1128,
  \dodoi{10.1088/0004-637X/718/2/1128}

\bibitem[{{Majewski} \& {Cudworth}(1993)}]{majewski1993}
{Majewski}, S.~R., \& {Cudworth}, K.~M. 1993, \pasp, 105, 987,
  \dodoi{10.1086/133269}

\bibitem[{{Malhan} \& {Ibata}(2018)}]{malhanibata2018arXiv}
{Malhan}, K., \& {Ibata}, R.~A. 2018, ArXiv e-prints, arXiv:1807.05994.
\newblock \doarXiv{1807.05994}

\bibitem[{McMillan(2011)}]{mcmillan2011}
McMillan, P.~J. 2011, \mnras, 414, 2446

\bibitem[{{Minniti} {et~al.}(2017){Minniti}, {Palma}, {D{\'e}k{\'a}ny},
  {Hempel}, {Rejkuba}, {Pullen}, {Alonso-Garc{\'{\i}}a}, {Barb{\'a}}, {Barbuy},
  {Bica}, {Bonatto}, {Borissova}, {Catelan}, {Carballo-Bello}, {Chene},
  {Clari{\'a}}, {Cohen}, {Contreras Ramos}, {Dias}, {Emerson}, {Froebrich},
  {Buckner}, {Geisler}, {Gonzalez}, {Gran}, {Hajdu}, {Irwin}, {Ivanov},
  {Kurtev}, {Lucas}, {Majaess}, {Mauro}, {Moni-Bidin}, {Navarrete},
  {Ram{\'{\i}}rez Alegr{\'{\i}}a}, {Saito}, {Valenti}, \&
  {Zoccali}}]{Minniti2017ApJ}
{Minniti}, D., {Palma}, T., {D{\'e}k{\'a}ny}, I., {et~al.} 2017, \apjl, 838,
  L14, \dodoi{10.3847/2041-8213/838/1/L14}

\bibitem[{{Monari} {et~al.}(2018){Monari}, {Famaey}, {Carrillo}, {Piffl},
  {Steinmetz}, {Wyse}, {Anders}, {Chiappini}, \& {Jan{\ss}en}}]{monari2018AA}
{Monari}, G., {Famaey}, B., {Carrillo}, I., {et~al.} 2018, \aap, 616, L9,
  \dodoi{10.1051/0004-6361/201833748}

\bibitem[{Myeong {et~al.}(2018)Myeong, Evans, Belokurov, Sanders, \&
  Koposov}]{Myeongetal2018ApJL}
Myeong, G.~C., Evans, N.~W., Belokurov, V., Sanders, J.~L., \& Koposov, S.~E.
  2018, \apjl, 863, L28, \dodoi{10.3847/2041-8213/aad7f7}

\bibitem[{Neuwirth(2014)}]{rcolorbrewer}
Neuwirth, E. 2014, RColorBrewer: ColorBrewer Palettes

\bibitem[{{Patel} {et~al.}(2017){Patel}, {Besla}, \& {Mandel}}]{patel2017MNRAS}
{Patel}, E., {Besla}, G., \& {Mandel}, K. 2017, \mnras, 468, 3428,
  \dodoi{10.1093/mnras/stx698}

\bibitem[{Plummer {et~al.}(2006{\natexlab{a}})Plummer, Best, Cowles, \&
  Vines}]{coda}
Plummer, M., Best, N., Cowles, K., \& Vines, K. 2006{\natexlab{a}}, R News, 6,
  7

\bibitem[{Plummer {et~al.}(2006{\natexlab{b}})Plummer, Best, Cowles, \&
  Vines}]{codapackage}
---. 2006{\natexlab{b}}, CODA: Convergence Diagnosis and Output Analysis for
  MCMC

\bibitem[{{Posti} \& {Helmi}(2019)}]{2019PostiHelmiAA}
{Posti}, L., \& {Helmi}, A. 2019, \aap, 621, A56,
  \dodoi{10.1051/0004-6361/201833355}

\bibitem[{{R Development Core Team}(2012)}]{R}
{R Development Core Team}. 2012, R: A Language and Environment for Statistical
  Computing, R Foundation for Statistical Computing, Vienna, Austria

\bibitem[{Ritchie {et~al.}(2015)Ritchie, Phipson, Wu, Hu, Law, Shi, \&
  Smyth}]{limmapackage}
Ritchie, M.~E., Phipson, B., Wu, D., {et~al.} 2015, Nucleic Acids Research, 43,
  e47

\bibitem[{{Rossi} {et~al.}(2015){Rossi}, {Ortolani}, {Barbuy}, {Bica}, \&
  {Bonfanti}}]{rossi2015}
{Rossi}, L.~J., {Ortolani}, S., {Barbuy}, B., {Bica}, E., \& {Bonfanti}, A.
  2015, \mnras, 450, 3270, \dodoi{10.1093/mnras/stv748}

\bibitem[{{Sakamoto} {et~al.}(2003){Sakamoto}, {Chiba}, \&
  {Beers}}]{sakamoto2003}
{Sakamoto}, T., {Chiba}, M., \& {Beers}, T.~C. 2003, \aap, 397, 899,
  \dodoi{10.1051/0004-6361:20021499}

\bibitem[{{Sohn} {et~al.}(2018{\natexlab{a}}){Sohn}, {van der Marel}, {Deason},
  {Bellini}, {Besla}, \& {Watkins}}]{sohn2018IAUS}
{Sohn}, S.~T., {van der Marel}, R.~P., {Deason}, A., {et~al.}
  2018{\natexlab{a}}, in IAU Symposium, Vol. 334, Rediscovering Our Galaxy, ed.
  C.~{Chiappini}, I.~{Minchev}, E.~{Starkenburg}, \& M.~{Valentini}, (Postdam,
  Germany), 47--50

\bibitem[{{Sohn} {et~al.}(2018{\natexlab{b}}){Sohn}, {Watkins}, {Fardal}, {van
  der Marel}, {Deason}, {Besla}, \& {Bellini}}]{2018SohnApJ}
{Sohn}, S.~T., {Watkins}, L.~L., {Fardal}, M.~A., {et~al.} 2018{\natexlab{b}},
  \apj, 862, 52, \dodoi{10.3847/1538-4357/aacd0b}

\bibitem[{Tierney {et~al.}(2013)Tierney, Rossini, Li, \& Sevcikova}]{snow}
Tierney, L., Rossini, A.~J., Li, N., \& Sevcikova, H. 2013, snow: Simple
  Network of Workstations

\bibitem[{{Vasiliev}(2019)}]{Vasiliev2019MNRAS}
{Vasiliev}, E. 2019, \mnras, 484, 2832, \dodoi{10.1093/mnras/stz171}

\bibitem[{Venables \& Ripley(2002)}]{mass}
Venables, W.~N., \& Ripley, B.~D. 2002, Modern Applied Statistics with S, 4th
  edn. (New York: Springer).
\newblock \url{http://www.stats.ox.ac.uk/pub/MASS4}

\bibitem[{Watkins {et~al.}(2010)Watkins, Evans, \& An}]{watkins2010}
Watkins, L., Evans, N., \& An, J. 2010, \mnras, 406, 264

\bibitem[{Watkins {et~al.}(2019)Watkins, van~der Marel, Sohn, \&
  Evans}]{Watkins2019ApJ}
Watkins, L.~L., van~der Marel, R.~P., Sohn, S.~T., \& Evans, N.~W. 2019, The
  Astrophysical Journal, 873, 118, \dodoi{10.3847/1538-4357/ab089f}

\bibitem[{{Wegg} {et~al.}(2019){Wegg}, {Gerhard}, \& {Bieth}}]{Wegg2019MNRAS}
{Wegg}, C., {Gerhard}, O., \& {Bieth}, M. 2019, \mnras, 485, 3296,
  \dodoi{10.1093/mnras/stz572}

\bibitem[{Wickham(2016)}]{hadleyGGPLOT}
Wickham, H. 2016, ggplot2: Elegant Graphics for Data Analysis (Springer-Verlag
  New York).
\newblock \url{http://ggplot2.org}

\bibitem[{{Wilkinson} \& {Evans}(1999)}]{wilkinsonevans1999}
{Wilkinson}, M.~I., \& {Evans}, N.~W. 1999, \mnras, 310, 645,
  \dodoi{10.1046/j.1365-8711.1999.02964.x}

\bibitem[{Xue {et~al.}(2008)Xue, Rix, Zhao, Fiorentin, Naab, Steinmetz, van~den
  Bosch, Beers, Lee, Bell, Rockosi, Yanny, Newberg, Wilhelm, Kang, Smith, \&
  Schneider}]{xue2008}
Xue, X.~X., Rix, H.~W., Zhao, G., {et~al.} 2008, \apj, 684, 1143,
  \dodoi{10.1086/589500}

\bibitem[{{Zoccali} {et~al.}(2001){Zoccali}, {Renzini}, {Ortolani}, {Bica}, \&
  {Barbuy}}]{zoccali2001}
{Zoccali}, M., {Renzini}, A., {Ortolani}, S., {Bica}, E., \& {Barbuy}, B. 2001,
  \aj, 121, 2638, \dodoi{10.1086/320411}

\end{thebibliography}

\end{document}